\documentclass[iop]{emulateapj}
\usepackage{epsfig}
\usepackage{amsmath,amssymb}
\usepackage{mathrsfs}
\usepackage{graphicx}
\usepackage{amsfonts}
\usepackage{bigstrut,array}

\begin{document}

\newcommand{\kms}{\mbox{ km s$^{-1}$}}
\newcommand{\be}{\begin{equation}}
\newcommand{\ee}{\end{equation}}

\def\ltsima{$\; \buildrel < \over \sim \;$}
\def\simlt{\lower.5ex\hbox{\ltsima}}
\def\gtsima{$\; \buildrel > \over \sim \;$}
\def\simgt{\lower.5ex\hbox{\gtsima}}
\newcommand\sgra{Sgr~A$^*$}
\newcommand\medd{\dot{M}_{\rm Edd}}
\newcommand\mcapt{\dot{M}_{\rm capt}}
\newcommand\ledd{{L}_{\rm Edd}}
\newcommand\mdot{\dot{m}}
\newcommand\Mdot{\dot{M}}
\def\del#1{{}}
\def\degs{$^\circ $}
\def\msun{{\,{\rm M}_\odot}}
\newcommand\mbh{{\,{\rm M}_{\rm bh}}}
\def\rsun{{\,R_\odot}}
\def\lsun{{\,L_\odot}}

\title{Supernovae in the
  Central Parsec: A Mechanism for Producing Spatially Anisotropic
  Hypervelocity Stars}

\author{Kastytis~Zubovas\altaffilmark{1}$^,$\altaffilmark{2}, Graham A. Wynn\altaffilmark{1} and Alessia Gualandris\altaffilmark{1}}

\altaffiltext{1} {Theoretical Astrophysics Group, University of
Leicester, Leicester LE1 7RH, U.K.}
\altaffiltext{2} {Center for Physical Sciences and Technology, 
Savanori\c{u} 231, Vilnius LT-02300, Lithuania; kastytis.zubovas@ftmc.lt}

\begin{abstract}
Several tens of hyper-velocity stars (HVSs) have been discovered escaping our
Galaxy. These stars share a common origin in the Galactic centre and are
distributed anisotropically in Galactic longitude and latitude.  We examine
the possibility that HVSs may be created as the result of supernovae occurring
within binary systems in a disc of stars around \sgra \ over the last 100 Myr.
Monte Carlo simulations show that the rate of binary disruption is $\sim
10^{-4}$~yr$^{-1}$, comparable to that of tidal disruption models. The
supernova-induced HVS production rate ($\Gamma_{\rm HVS}$) is significantly
increased if the binaries are hardened via migration through a gaseous
disc. Moderate hardening gives $\Gamma_{\rm HVS} \simeq 2 \times
10^{-7}$~yr$^{-1}$ and an estimated population of $\sim 20$ HVSs in the last
$100$~Myr. Supernova-induced HVS production requires the internal and external
orbital velocity vectors of the secondary binary component to be aligned when
the binary is disrupted. This leaves an imprint of the disc geometry on the
spatial distribution of the HVSs, producing a distinct anisotropy.

\end{abstract}


\section{Introduction} \label{sec:intro}

In recent years, several tens of stars have been discovered moving
through the Galaxy with large radial velocities
\citep{Brown2005ApJ,Brown2007ApJ}. They are usually called
`hypervelocity stars' (HVSs), although the term is not precisely
defined. \citet{Brown2012ApJ} define HVS as a star with a Galactic
rest-frame radial velocity $v_{\rm rf} > 275$~km/s, with a
sub-population of unbound HVSs having $v_{\rm rf} > 400$~km/s.

Most HVSs are B-type stars with ages $t_{\rm HVS} \lesssim 200$ Myr
\citep{Brown2012ApJb}. The difference between the stellar age and the
travel time from the Galactic centre (called the ``arrival time'') has
been determined for a few HVSs and lies between several tens - 100 Myr
\citep{Brown2012ApJb}.

The HVSs follow a statistically significant anisotropic distribution
in galactic longitude and latitude and seem to form an extended, thick
disc \citep{Brown2009ApJ,Brown2012ApJ}. HVS velocities and orbits
\citep[e.g.][]{Brown2005ApJ} strongly suggest an origin in the
Galactic Centre (GC) close to \sgra. Few proper motion measurements of
HVSs exist; two HVSs seem to originate in the Galactic disc
\citep{Heber2008A&A, Tillich+2009} and one seems to be confirmed as
ejected from the GC \citep[but see \citealt{Irrgang2013A&A} for a
  claim that the star's origin in the LMC cannot be ruled
  out]{Brown2010ApJ}.

The dominant model explaining HVS ejection is the tidal disruption of
binaries by \sgra, the supermassive black hole (SMBH) at the centre of
our Galaxy. This model, and associated assumptions, can explain the
production rates and velocities of the observed HVS population. An
alternative model with a Galactic disc origin, the disruption of
binary stars by supernova explosions, does not produce large enough
velocities. We discuss HVS production mechanisms in more detail in
Section \ref{sec:prodmech} below.  We note that there is another
definition of HVS based on their production mechanism, with stars
created via tidal disruption of binary stars (see below) termed HVSs
and those produced by other channels termed `runaway' or
`hyper-runaway' stars. In this Paper, we define HVSs only by their
radial velocities.

In this paper we develop the idea that HVSs can be produced via
supernova-induced disruption of binaries in the GC
\citep{Baruteau2011ApJ}. We show this model to be consistent with the
GC origin, anisotropic spatial distribution and arrival times of the
observed HVS population. We propose that a disc of stars analogous to
the discs of young stars currently observed in the central parsec
\citep{Paumard2006ApJ} formed during extended GC star formation over
the last $\sim 10^8$~yr \citep{Blum2003ApJ}. Core-collapse
  supernovae will have disrupted many of these binaries, ejecting
some secondary (less massive) stars with a range of terminal
velocities. When the orbital velocities of binaries around \sgra
\ align with the internal orbital velocities of secondary stars within
the binaries, the terminal velocity can reach as much as $>600$~km/s,
potentially explaining most of the known HVSs. The original stellar
disc provides a natural plane for the HVS spatial distribution. We
find HVS production rates comparable to tidal disruption models and
arrival times consistent with observation.

We begin by outlining the commonly investigated HVS production
scenarios and highlight their major features in Section
\ref{sec:prodmech}. The physical basis for the model and simulations
is presented in Section \ref{sec:analyt}. Section \ref{sec:results}
presents the expected HVS production rates, spatial distribution and
arrival times.  We discuss the results in Section \ref{sec:discuss}
and conclude in Section \ref{sec:concl}.

\section{HVS production mechanisms} \label{sec:prodmech}

The earliest and most popular mechanism proposed for the ejection of
HVSs is the tidal disruption of stellar binaries by the SMBH in the
Galactic Center. In this scenario, originally proposed by
\citet{Hills98} and later studied by \citet{Yu2003ApJ}, stellar
binaries on low angular momentum orbits come close enough to the SMBH
to be tidally disrupted in a three-body interaction. As a result, one
star acquires a large velocity, possibly becoming an HVS, while the
companion remains bound to the SMBH in a very eccentric orbit. Such
bound stars may represent the progenitors of the S-stars
\citep[e.g.][]{Gillessen+2009,Perets+2009}. Models show that the
  S-stars and the HVSs can be produced from the same parent population
  \citep{Zhang2013ApJ}. Because of the large abundance of binaries in
dense stellar environments, this process is considered a natural
occurrence in galactic nuclei hosting supermassive black holes.  The
discovery of the first HVS star \citep{Brown2005ApJ} helped to
validate the scenario.  The tidal disruption model predicts a
continuous ejection of stars from the Galactic Center and an isotropic
distribution of HVSs in the halo, unless the original stellar binaries
are prevalently found in flattened systems like stellar disks
\citep{Lu2010ApJ}.  In the presence of a binary black hole, encounters
with single stars also result in the acceleration of stars to
hypervelocities, with a somewhat higher rate.  \citet{Yu2003ApJ}
predict an ejection rate of the order of $10^{-5}$~yr$^{-1}$ for a
single SMBH and $10^{-4}$~yr$^{-1}$ for a black hole binary, in
agreement with available observations.  However, production rates
depend on several assumptions about the binary fraction in the
Galactic Center, the state of the loss cone and the number of massive
perturbers, such as giant molecular clouds or star clusters
\citep{Perets2007ApJ}, and are uncertain within two or three orders of
magnitude.  Possible difficulties of the tidal disruption model
include the observed anisotropic distribution of HVSs in the halo and
the existence of HVSs with measured proper motion not originating from
the Galactic Center \citep{Heber2008A&A,Tillich+2009}.
 
High velocity stars, generally called runaways, are produced in the
Galactic disk by the disruption of massive binaries due to either
supernova explosions or dynamical encounters.  In the binary supernova
mechanism, a stellar binary is unbound when the primary star explodes
as supernova \citep{Blaauw1961BAN}, due to mass loss and a natal kick
if the explosion is asymmetric.  In this case, the companion star is
ejected with a velocity approximately equal to the binary orbital
velocity \citep{Hills1983ApJ}, i.e. no more than $\sim 200-300\kms$
\citep[e.g.][]{SPZ2000, Dray2005,GPZS2005}.  In the dynamical ejection
mechanism, stars are ejected as a result of close encounters between
stars and binaries in dense stellar regions \citep{Leonard1991}, and
ejection velocities can be somewhat higher, being limited by the
escape velocity of the most massive star. In addition, if runaways are
ejected in the direction of Galactic rotation, their velocity is
incremented by the orbital speed at their location. In both scenarios,
runaways tend to be concentrated close to the disk plane, with typical
vertical height dependent on mass \citep{Bromley2009ApJ}.  While most
runaway stars are bound to the Galaxy, a few have been discovered to
be unbound \citep{Przybilla2008ApJ, Irrgang2010}, and hence named
hyper-runaways, though some argue they should be considered HVSs.  The
production of unbound runaway stars has been confirmed by numerical
simulations \citep{Gvaramadze2009MNRAS, GG2011}.

An alternative mechanism for the production of HVSs is the tidal
disruption of dwarf galaxies in the galactic potential
\citep{Abadi+2009}, though the contribution to the observed sample may
be small \citep{Piffl+2011}.

\section{Supernovae in Galactic centre binaries} \label{sec:analyt}

\subsection{Supernova rates}

\begin{figure}
  \centering
    \includegraphics[width=0.45\textwidth]{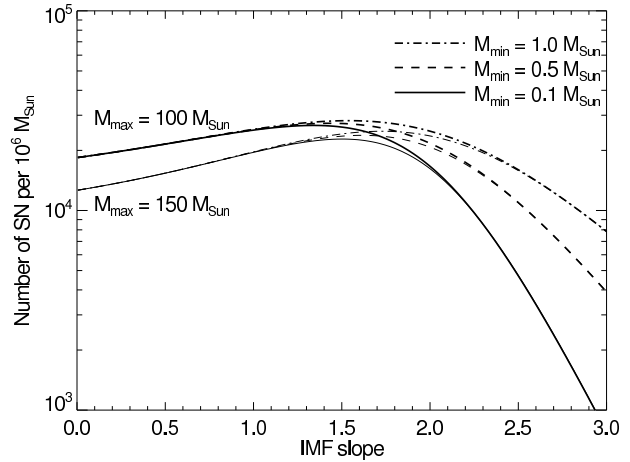}
  \caption{Number of supernovae in $10^8$~yr per $10^6 \; \msun$ of
    initial stellar mass and its dependence on the IMF power-law slope
    $\Gamma$ and the minimum and maximum stellar mass.}
  \label{fig:nsn}
\end{figure}

Figure \ref{fig:nsn} shows the expected number of supernovae (SNe) within
$10^8$ years of a star formation event per $10^6 \; \msun$ of initial stellar
mass and its dependence on the IMF power-law index $\Gamma$ (where ${\rm
  d}N(M)/{\rm d}M \propto M^{-\Gamma}$) and the stellar mass limits. The
number of SNe is assumed to be equal to the number of stars more massive than
8$\msun$, the supernova limit \citep[by $100$~Myr all of these stars will have
  undergone supernovae;][]{Woosley2002RvMP}. Observations
\citep{Kollmeier2009ApJ} and theory \citep{Nayakshin2006MNRASb} constrain the
IMF of stars close to \sgra \ to be normal or top-heavy (i.e. $\Gamma >
-2.3$), so that we expect $\sim 1-2 \times 10^4$ SNe per $10^6 \; \msun$ of
stars formed in the Galactic centre 100~Myr ago.

The star formation history of the GC is poorly constrained, but observations
\citep{Blum2003ApJ} indicate that $\gtrsim 3\times 10^5 \; \msun$ of stars
formed within $r = 2.5$~pc of \sgra \ in the last $10^8$~yr and a similar or
higher mass formed between $1$~Gyr -- $100$~Myr ago.  We investigate initial
stellar masses in the range $\sim 3-10 \times 10^5 \;\msun$ with a high binary
fraction $f_{\rm bin} = 0.7-1$ \citep[e.g.,][]{Kobulnicky2007ApJ}, giving GC
supernova rates $\sim 2 \times 10^{-5} - 2 \times 10^{-4}$~yr$^{-1}$. We use a
fiducial value of $10^{-4}$~yr$^{-1}$ when presenting results.

\subsection{Effects of a supernova kick}

Observations of isolated radio pulsars \citep{Hobbs2005MNRAS,
  Arzoumanian2002ApJ} suggest that many core collapse SNe produce remnants
with kick velocities
\begin{equation}
30 \: {\rm km \: s^{-1}} \lesssim v_{\rm k} \lesssim 1200 \: {\rm km \: s^{-1}} 
\end{equation}
(corresponding to $1\sigma$ deviations from the two peaks in the distribution
found by \citealt{Arzoumanian2002ApJ}).  For binaries
with equal mass, $10 \msun$ components and typical orbital periods $1 - 300$
days \citep{Raguzova2005A&AT} the orbital velocities of the members around
their common centre of mass (the ``internal'' orbit) range between
\begin{equation}
50 \: {\rm km \: s^{-1}}  \lesssim v_{\rm int} \lesssim 320 \: {\rm km \: s^{-1}} . 
\end{equation}
Clearly, the kick velocities can be sufficient to unbind massive binaries in
this period range.

The velocities of binaries in circular orbits around \sgra \ (the ``external''
orbit) follow
\begin{equation}
v_{\rm ext} \simeq 130 R_{\rm pc}^{-1/2} \: {\rm km \: s^{-1}} , 
\end{equation}
where $R_{\rm pc}$ is the radius of the external orbit in parsecs.  These
external orbital velocities are of comparable magnitude to the internal
orbital velocities and their vector sum determines the subsequent orbits of
ejected binary members. If the internal and external orbital velocity vectors
of a secondary star are aligned when the binary is disrupted, then the
secondary may be ejected from the GC as an HVS.  The natal kick helps unbind
the binary, but it is the combination of the internal and external orbital
velocities that produces HVSs. This alignment requirement naturally results in
an anisotropic distribution of HVSs, with a higher density of objects in the
plane of the stellar disc. Since the disc alignment is unknown (the
present-day discs are not aligned with any larger structures in the Galaxy),
the observed plane in the HVS distribution could represent a fossil imprint of
the orientation of a $\lesssim 10^6 \; \msun$ stellar disc in the GC $\sim
100$ Myr ago.

\subsection{Model description}

We use Monte Carlo simulations of binary supernovae in the GC to
constrain HVS production rates and possible orientations of the proposed 
stellar disc. 

Binaries are initially distributed in a thin disc (external orbits at $z = 0$)
extending from $r_{\rm in} = 0.04$~pc to $r_{\rm out} = 2.5$~pc from a $4
\times 10^6 \: M_\odot$ point mass. The radial stellar density profile is a
power law with $\Sigma \propto r^{-0.5}$. Calculations assume circular,
Keplerian external orbits, appropriate for $r_{\rm out} \lesssim
3$~pc. Primary masses range between $M_{\rm p,min}=8$ and $M_{\rm p,max} =
100\:\msun$ and follow a Salpeter IMF. Secondary masses are selected from
$M_{\rm s,min} = 1 \: \msun - M_{\rm p}$.  Binary mass ratios are uniformly
distributed.

Following \citet{Sana2012Sci}, binary internal orbital periods are
selected from the range $1 - 300$ days, following a $({\rm log}P/{\rm
  days})^{-0.55}$ distribution. Internal orbital planes are randomly
orientated and uncorrelated with the disc plane.  We also consider the
effects of binary hardening (evolution to shorter internal orbital
periods), which may happen in a few tens of external orbits as
binaries migrate through a gaseous disc \citep{Baruteau2011ApJ}.  We
examine the cases of no hardening, weak hardening (binary separation
reduced by 2) and strong hardening (separation reduced by 10), with
circular orbits assumed in all cases.  In each case, we reject binary
systems which, after hardening, would have separation smaller than the
sum of the two stellar radii. The latter are calculated using equation
(2) in \citet{Gvaramadze2009MNRAS}.

Pre-supernova evolution is governed by mass loss from the primary. We adopt
the pre-explosion -- ZAMS mass relation of \citet{Limongi2009MmSAI}. Supernova
mass loss is assumed to be instantaneous and remnant masses are taken from
\citet{Limongi2009MmSAI}. Kick velocities are drawn from the bimodal
distribution of \citet{Arzoumanian2002ApJ} and follow an isotropic spatial
distribution.  These are added to the remnant orbital velocities.  We do not
consider the effects of the supernova ejecta on the secondary.

For disrupted binaries, we calculate the velocities of the secondaries after
they escape the remnant. For those secondaries with velocities greater than
the escape velocity from the SMBH, we calculate their terminal
velocities. (Real velocities would be somewhat smaller since we do not
consider the extended background potential.)  If the terminal velocity is
greater than $500$~km/s, we label the star as an HVS.

Each calculation samples $N_{\rm tot} = 10^7$ SNe for each case of binary
hardening. The results were divided into 20 subsamples to examine statistical
variations. Statistical convergence was tested by varying $N_{\rm tot}$, with
$N_{\rm tot} = 10^7$ being well converged.

\section{Results} \label{sec:results}

\begin{table*}
  \centering
  \caption{Probabilities of binary disruption and HVS ejection by supernova explosions}
  \setlength{\extrarowheight}{1.5pt}
  \begin{tabular}{l c | c c c c}
    \hline
    \hline
    Model & Hardening$^a$ & $p_{\rm disr}$ $\!^b$ & $p_{\rm unb}$ $\!^c$ & $p_{\rm 275}$ $\!^d$ & $p_{\rm 400}$ $\!^e$ \bigstrut\\
    \hline
    H1     & None   & $96.6 \pm 0.1 \%$  & $3.92 \pm 0.15 \%$ & $0.37 \pm 0.10 \%$ & $0.06 \pm 0.03 \%$ \\
    \hline
    H0.5   & Weak   & $95.5 \pm 0.1 \%$  & $7.26 \pm 0.11 \%$ & $0.75 \pm 0.09 \%$ & $0.19 \pm 0.04 \%$ \\
    \hline
    H0.1   & Strong & $93.5 \pm 0.1 \%$  & $21.66 \pm 0.28 \%$ & $5.12 \pm 0.18 \%$ & $1.09 \pm 0.08 \%$ \\
   \hline
    \hline
  \end{tabular}
  \begin{list}{}{}
  \item[ ] {\footnotesize $^a$ ``Hardening" refers to a reduction in binary
    separation caused by migration (see text);

$^b$ probability that a supernova disrupts a binary;

$^c$ probability that a secondary star becomes unbound from the SMBH;

$^d$ probability that a secondary has terminal velocity $> 275$~km/s;

$^e$ probability that a secondary has terminal velocity $> 400$~km/s.}

  \end{list}

  \label{table:results}
\end{table*}

Table \ref{table:results} presents the results of the Monte Carlo
calculations. The models differ in the extent of binary hardening.  
All production probabilities are given per binary supernova.

\subsection{Binary disruption rate}

The probability that a supernova disrupts a binary is $p_{\rm disr} \gtrsim 93
\%$. This result is higher than earlier models \citep[e.g.,][where $p_{\rm
    disr} \sim 70-80 \%$]{Hills1983ApJ} because we consider SN progenitors
with masses down to $8 \; \msun$. Low-mass progenitors lose a larger fraction
of their mass during the explosion and so are more easily unbound. Hence,
top-heavy IMFs give lower binary disruption rates. We ran model H0.5 with an
IMF slope of 1.3 and found $p_{\rm disr} = 90\%$, with a negligible effect on
the other probabilities (see below).  Disruption rates are independent of
external orbital radii.

The high disruption probabilities yield a fiducial supernova-induced binary
disruption rate for the GC $\Gamma_{\rm disr} \sim 10^{-4}$~yr$^{-1}$, with
uncertainties giving a range $2 \times 10^{-5} - 2 \times
10^{-4}$~yr$^{-1}$. This is comparable to the tidal disruption rates of
binaries as found by \citet{Perets2007ApJ}. In that model, tidal disruption
rates are enhanced by rare encounters with massive perturbers, such as giant
molecular clouds.  We find similar disruption rates as a natural result of SNe
in massive binaries without recourse to external encounters.

\subsection{Stellar ejection rate}

\begin{figure}
  \centering
    \includegraphics[width=0.45\textwidth]{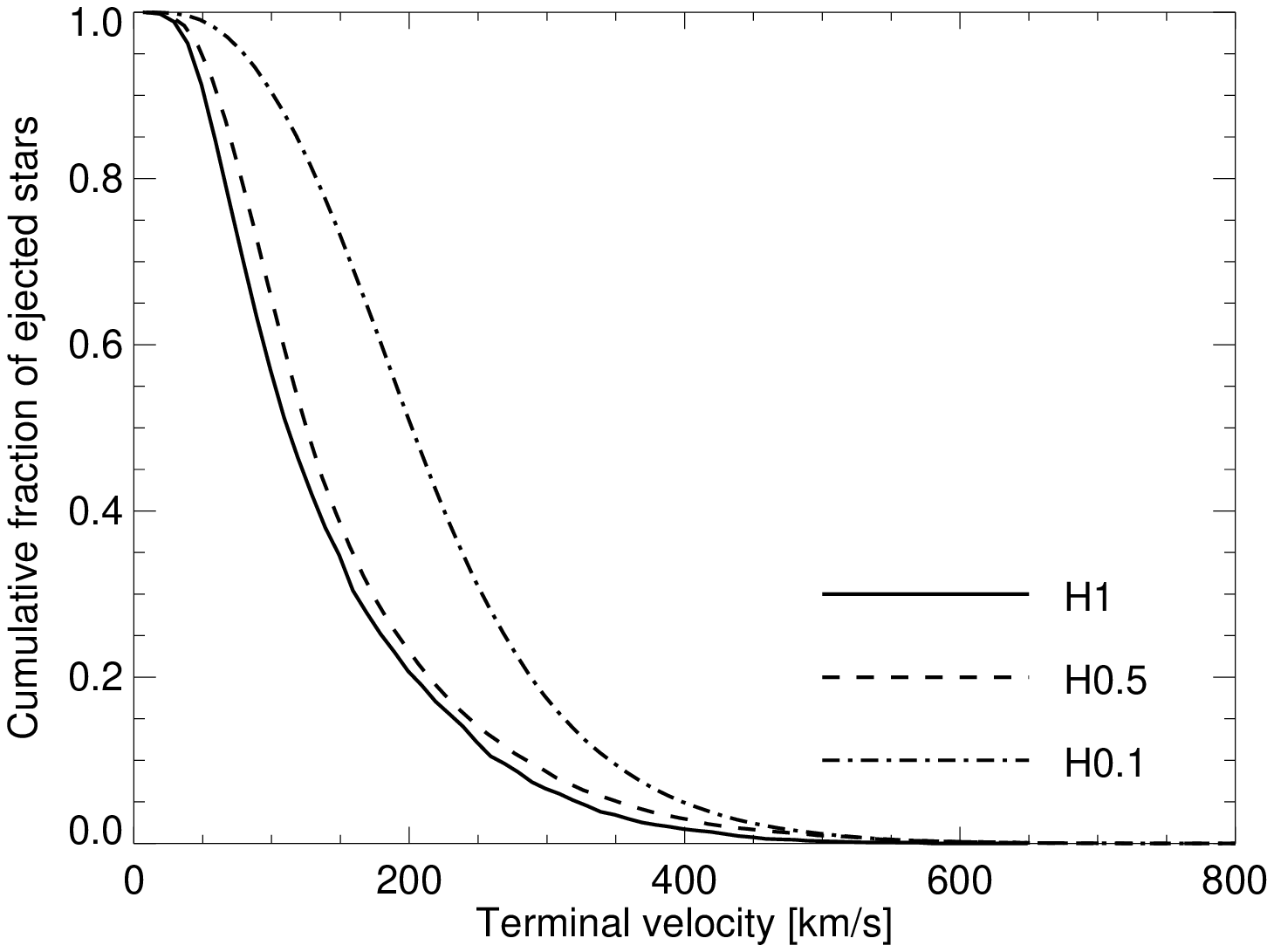}
    \includegraphics[width=0.45\textwidth]{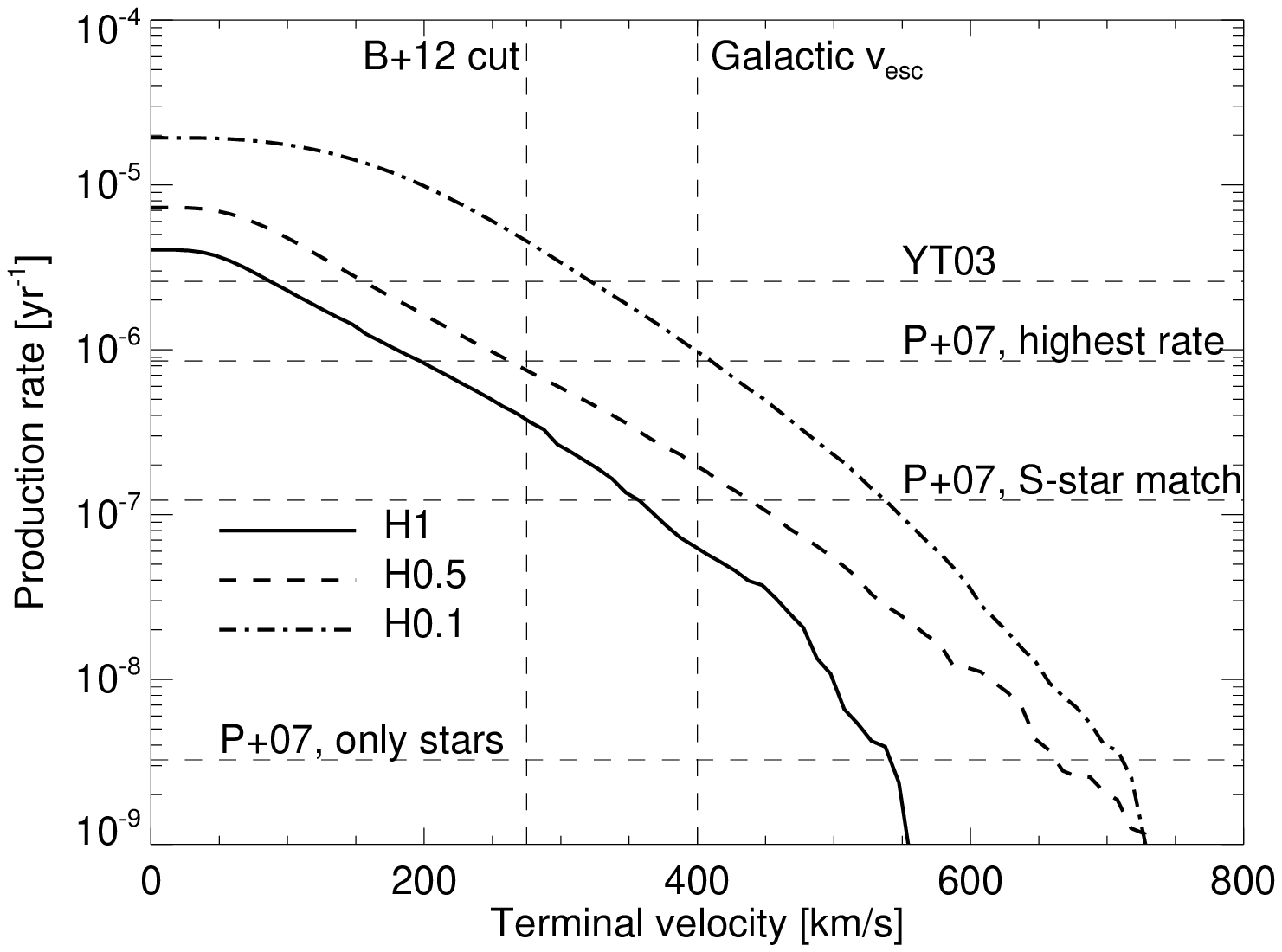}
  \caption{{\bf Top panel:} distribution of ejected
    star velocities as a fraction of all ejected stars for the three
    models. {\bf Bottom panel:} Production rates of stars as a function of
    terminal velocity; three thick lines correspond to the three
    models presented in this paper. Horizontal dashed lines: HVS
    production rates for tidal disruption models, see text for
    details. Vertical dashed lines: velocity cuts from
    \citet{Brown2012ApJ}.}
  \label{fig:termvel}
\end{figure}

\begin{figure*}
  \centering
    \includegraphics[width=1.00\textwidth]{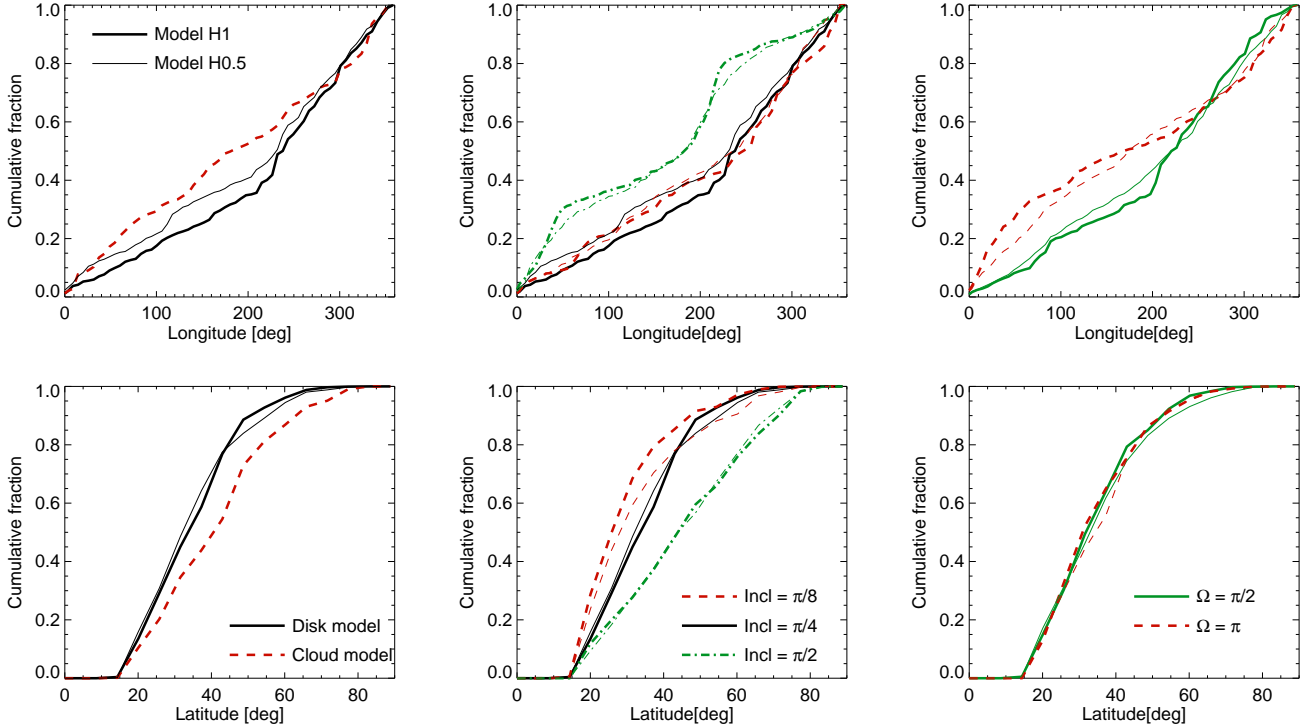}
  \caption{Variation of final HVS spatial distribution for models H1 (thick
    lines) and H0.5 (thin lines) with initial binary spatial distribution. We
    neglect all stars that would be hidden by the Galactic plane,
    ($\left|b\right|<20^\circ$) and reduce statistical weight of stars with $b
    < 0$ by 3 to account for lack of southern hemisphere observations in
    \citet{Brown2012ApJ} (see their figure 4). {\bf Top panels:} cumulative
    fraction of stars vs Galactic longitude. {\bf Bottom panels:} cumulative
    fraction vs Galactic latitude. {\bf Left:} comparison of a thin disc
    inclined by $45^\circ$ with line-of-nodes along $\Omega = 120^\circ$ with
    a spherically symmetric initial distribution. Disc geometry causes a clear
    anisotropy. {\bf Middle:} dependence on the inclination of the disc plane;
    modest inclinations do not affect the longitude distribution
    significantly, while a large inclination results in peaks at $\Omega \pm
    90^\circ$. {\bf Right:} dependence on the line-of-nodes position; the
    position of the longitude peak shifts while the latitude distribution is
    unaffected.}
  \label{fig:longspread}
\end{figure*}

\begin{figure}
  \centering
    \includegraphics[width=0.45\textwidth]{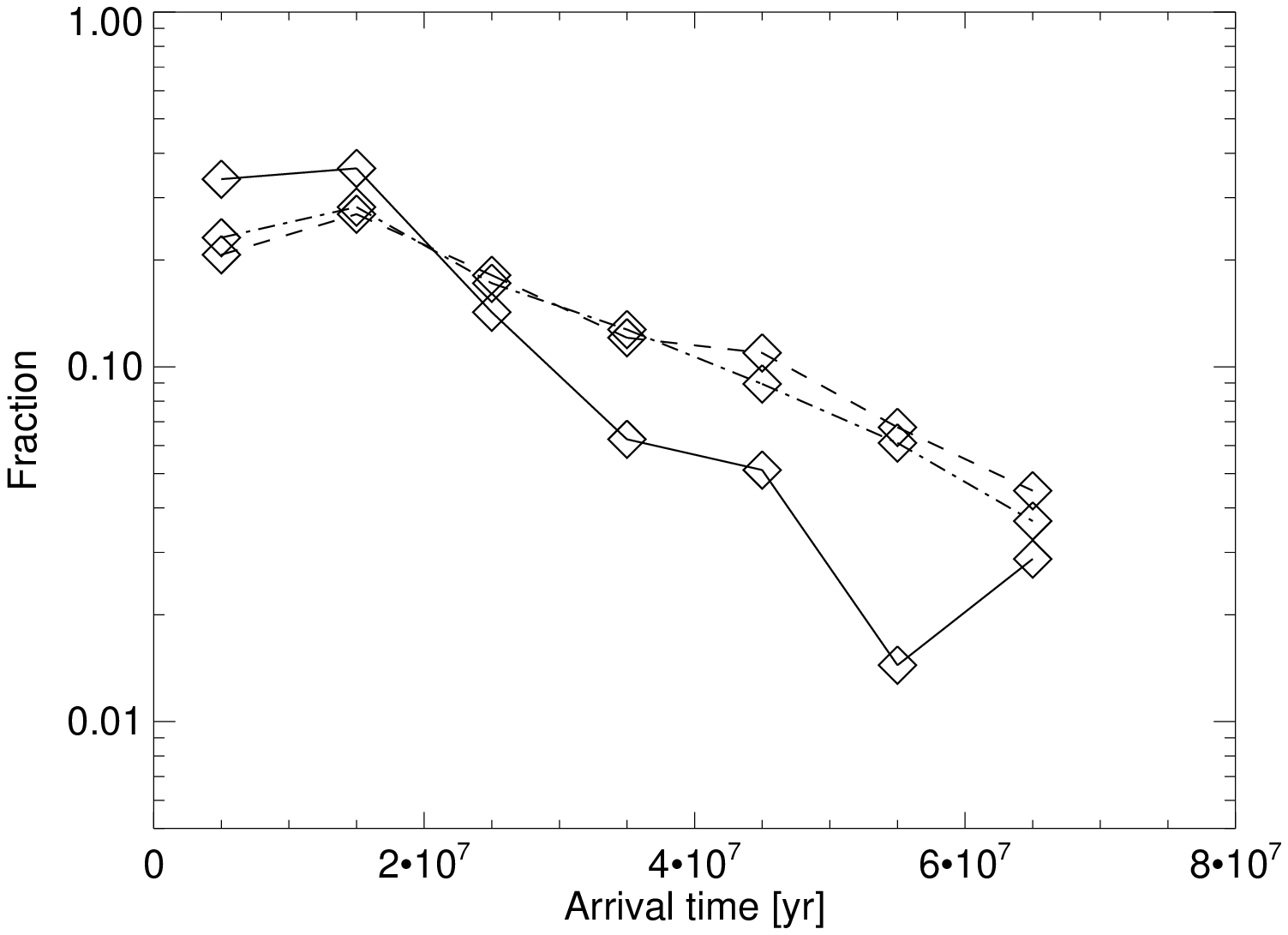}
  \caption{Fraction of HVSs as a function of the main sequence lifetime of the
    primary, binned by $10^7$~yr, for models H1 (solid), H0.5 (dashed) and
    H0.1 (dot-dashed line). The lowest mass supernova progenitors spend $\sim
    70$~Myr on main sequence, potentially explaining the HVSs with arrival
    times close to $10^8$~yr.}
  \label{fig:lifetimes}
\end{figure}

Although most secondary stars remain bound to the SMBH even after
binary disruption, a fraction attains a large enough velocity to
escape from the sphere of influence. We give the secondary star
ejection probability per SN explosion in the fourth column of Table
\ref{table:results}; it ranges from $\sim 4 \%$ for unhardened
binaries to $\sim 25 \%$ for strong hardening. For moderate hardening
we find a stellar ejection rate of $\sim 8 \times 10^{-6}$~yr$^{-1}$.

The production rate of unbound stars is $\sim 3-4$ times higher for
binaries with external orbital radii $R \sim 0.1-0.5$~pc, where
$v_{\rm ext} \sim v_{\rm int}$, than elsehwere.  Closer to the SMBH,
$v_{\rm ext} \gg v_{\rm int}$ and the central potential is too deep
for stars to easily escape.  Further out, external velocities are too
low to significantly boost the terminal velocities.  Hence, our
results are insensitive to the inner and outer disc boundaries.

We present the cumulative distribution of ejected star terminal
velocities in Figure \ref{fig:termvel}. In the top panel, we show the
distributions as fractions of all ejected stars. This shows that
models with stronger hardening not only eject more stars, but also
eject them with higher average velocities. In the bottom panel, the
distributions are convolved with the supernova rate of
$10^{-4}$~yr$^{-1}$ and the probabilities of stellar ejection. We also
show the HVS production rates predicted by tidal disruption models:
\citet{Yu2003ApJ}, who considered binary injection into low angular
momentum orbits due to stellar relaxation only (dashed line labelled
`YT03'); an equivalent model by \citet{Perets2007ApJ}, who used a
better-constrained binary distribution to find a production rate a
factor $10^3$ lower (`P+07, only stars'); and two models from
\citet{Perets2007ApJ} involving relaxation accelerated by
gravitational interactions with massive perturbers such as GMCs
(`P+07, highest rate' refers to their model GMC1, which gives the
highest binary disruption rate and largest number of HVSs, while
`P+07, S-star match' refers to model GMC2, which provides the number
of S-stars in the central $\sim 0.04$~pc around \sgra consistent with
observations). The two vertical dashed lines represent the velocity
cut at $275$~km/s used by \citet{Brown2012ApJ} to identify HVSs and
the $400$~km/s Galactic escape velocity used by the same
authors\footnote{The precise value of the Galactic escape
    velocity depends on the assumed parameter of the dark matter halo
    and can be larger than $400$~km/s.}. We also give the production
probabilities per SN explosion for stars with velocities higher than
the two thresholds in Table \ref{table:results}.

In $10^8$ years, our models H1, H0.5 and H0.1 predict $37$, $75$ and
$512$ stars ejected with terminal velocities $v_{\rm term} >
275$~km/s, respectively. These values are somewhat optimistic, because
we do not consider deceleration while escaping the Galactic
gravitational potential. The latter can be crudely approximated as
$\Delta \Phi \sim \sigma^2 {\rm ln} \left(R/R_{\rm infl}\right)$, with
$\sigma = 100$~km/s the velocity dispersion in the Galaxy and $R_{\rm
  infl} \simeq 2$~pc the radius of influence of \sgra. Taking $R =
R_{\rm Solar} = 8$~kpc, we find that only stars with $v_{\rm
  term}\simgt 400$~km/s have radial velocities above the cut at the
Solar circle. For these, our models predict $6$, $19$ and $109$ such
stars. Therefore, a binary orbit hardening by slightly more than a
factor $2$ is enough to produce HVS numbers consistent with
observations. In addition, the ratio of unbound to bound HVSs is $\sim
0.3$ in all our models, slightly lower than the observationally
derived $\sim 0.5$ \citep{Brown2012ApJ}.

The highest velocities we expect to find in $10^8$~yr are $\sim500$,
$\sim600$ and $\sim700$~km/s for models H1, H0.5 and H0.1,
respectively. Accounting for the deceleration while escaping the
galaxy, these velocities are reduced to $\sim400$, $\sim530$ and
$\sim640$~km/s, respectively. Such velocities cannot explain the
fastest known HVSs, for example US708 with $v_{\rm rf} \simeq
  750$~km/s \citep{Hirsch2005A&A}, but the vast majority of currently
known ones are moving slower than these maximal radial velocities.

\subsection{Spatial distribution of HVSs}

\citet{Brown2012ApJ} present observational evidence for anisotropy in the
spatial distribution of HVSs (see their Figures 4 and 5). They find a strong
excess of HVSs in the Northern galactic hemisphere at longitudes $240^\circ <
l < 300^\circ$ and a mild excess at latitudes $45^\circ < b <
60^\circ$. Overall, their result provides a $3\sigma$ confirmation that the
HVS distribution is not spherically symmetric.

We produce plots equivalent to Figure 5 from \citet{Brown2012ApJ} in
Figure \ref{fig:longspread} for models H1 and H0.5. We apply a similar
a-priori cut as in \citet{Brown2012ApJ}: we only consider stars with
$\left|b\right|>20^\circ$ and reduce the statistical weight of stars
with $b < 0$ by a factor 3.  We consider only stars with terminal
velocity $v_{\rm term} > 400$~km/s; as we showed above, these stars
would have $v_{\rm radial} \sim 275$~km/s at the Solar circle,
equivalent to the velocity cut \citet{Brown2012ApJ} used to define
HVSs. We compare the HVS distribution obtained from an thin disc with
an inclination $i = 45^\circ$ to the Galactic plane and line-of-nodes
direction $\Omega$ pointing toward Galactic longitude $l = 120^\circ$
with an initially spherically symmetric distribution in the left
panels. We find a clear anisotropy caused by the disc. The middle
panels show the effects of disc inclination.  Angles $i \lesssim
60^\circ$ yield a longitude distribution similar to that observed,
while $i \simeq 45^\circ$ produces a comparable latitude
distribution. The effects of the line-of-nodes direction $\Omega$ are
shown in the right panels and the longitude distribution can be seen
to have a strong peak at $\Omega + \pi/2$ and a weaker one at $\Omega
- \pi/2$. The anisotropy is weaker in the H0.5 model than in H1
because hardened binaries have larger internal orbital velocities,
weakening the effect of correlated external motions (i.e. the
disc). Nevertheless, anisotropy is still present in H0.5. The
anisotropy would be amplified if the internal orbits aligned with the
disc due to the same torques that cause binary hardening.

\subsection{Arrival times}

\citet{Brown2012ApJb} recently put constraints on the difference between the
stellar age and the travel time from the Galactic centre for five HVSs (the
``arrival time''). In our model, arrival time is approximately equal to the
main sequence lifetime of the primary star. We plot the fraction of HVSs as a
function of the lifetime of the primary star in Figure
\ref{fig:lifetimes}. The main sequence lifetime is given by $t_{\rm MS} \simeq
1.2 \times 10^{10} \left(M / \msun\right)^{-2.5}$~yr and is a very steep
function of mass. Stars with $M \simeq 8 \; \msun$ survive for $\sim 70$~Myr
and will have been the companions of HVSs with the longest arrival times.

We find $\sim 5-10 \%$ of HVSs have companions with main sequence lifetimes
$t_{\rm MS} > 50$~Myr, within $3\sigma$ of the arrival time results of
\citet{Brown2012ApJb}. It it also worth noting that massive stars tend to have
companions of similar masses \citep{Halbwachs2003A&A}, such that the
companions of the most massive stars are more likely to have undergone SNe
explosions themselves, reducing the fraction of observable HVSs with short
arrival times.

\section{Discussion} \label{sec:discuss}

Our results suggest that supernova-induced disruption of binaries in
the Galactic centre can eject stars with rates high enough to explain
the majority of the observed HVSs, provided that binaries are hardened
compared to their counterparts in the field. We now discuss the
processes that potentially lead to binary hardening, the influence of
Type Ia supernovae on our results and the fate of undisrupted
binaries.

\subsection{Binary hardening}

\citet{Baruteau2011ApJ} present an analysis of binary evolution and
migration through gaseous discs. One of the findings is that the disc
exerts a torque on the migrating binaries, which hardens them. Typical
hardening timescales are a few tens of external orbital times.  In our
case, the external orbital timescale is $t_{\rm orb} \sim 7500 R_{\rm
  pc}^{3/2}$~yr and binaries are expected to harden on a timescale
$t_{\rm harden} \sim 10^5$~yr, much shorter than the lifetime of even
the most massive stars. This validates our assumption of binary orbits
remaining circular after hardening.

Since the hardening torque acts perpendicular to the disc plane, it
also works to align the binary internal orbit with its external orbit
within the disc. The effect is unlikely to be strong, as the torque
direction is essentially random on length scales much smaller than the
disc scale height, $H \sim 200 R_{\rm pc}$~AU for $H/R = 10^{-3}$, but
may lead to a weak correlation between internal and external orbital
velocity planes. Our calculations show that even a perfect correlation
does not increase the stellar ejection rates significantly, however
the spatial anisotropy of ejected star directions does become more
pronounced.

Another potential binary hardening mechanism is the common envelope
phase at the end of the primary's life. This phase of binary star
evolution may occur when the binary separation is of the same order as
the primary star's radius so that the secondary star is engulfed by
the envelope of the primary.  The binary orbit is then hardened as
(internal) orbital energy is used in unbinding the primary's
envelope. The efficiency with which orbital energy is extracted by
this process is very uncertain \citep[see e.g.\ ]{Ivanova2013A&ARv}
and we do not include its effects in our model. However, we note that
common envelope evolution would be more frequent in a binary
population with orbits hardened via migration through a disc and may
play an important role in HVS production.

\subsection{Type Ia supernovae}

In our calculations, we have considered only core-collapse supernova
explosions. White dwarfs in binary systems can explode as Type Ia
supernovae; since there is no remnant in that case, the binary
disappears and the secondary star can also be ejected from the GC. It
is difficult to estimate the rate of these supernovae, because it
depends on many parameters of the system, so we cannot constrain the
rate of ejection of low-mass stars. Their dynamics, however, are very
similar to those presented in this paper. Although the binary
components have lower masses, the orbital separations can also be
lower due to one binary component being compact, therefore the
internal orbital velocities should be comparable. Furthermore, since
the secondary star does not need to escape the gravity of the
supernova remnant, it loses less orbital energy while leaving the
system. \citet{Pan2013arXiv} recently interpreted the HVS US708
\citep{Hirsch2005A&A} as a possible former companion of a Type Ia SN
progenitor.

\subsection{Pulsars and XRBs in the Galactic Centre}

Undisrupted binaries may remain visible as pulsars or X-ray binaries until the
secondary star dies. Using an approximate $5 \%$ non-disruption probability
(see Table \ref{table:results}), we estimate a XRB creation rate $\Gamma_{\rm
  puls} < 10^{-5}$~yr. Typical XRB lifetimes $\lesssim 10^7$~yr
\citep{King2001ApJ} lead us to predict a population of $< 100$ Galactic centre
XRBs at any time. This number may be significantly reduced as undisrupted
binaries are likely to have eccentric orbits and possibly be prone to tidal
disruption. The fate of XRBs and the remnants in general is complicated by
stellar dynamics within the GC, binary evolution and accretion physics and we
intend to examine these issues in future work.

\section{Conclusions} \label{sec:concl}
 
We have used a Monte Carlo model to examine the effects of
core-collapse SN explosions in binary stars in orbit around \sgra. We
calculate the post-SN orbits of the binary members and find that most
binaries get disrupted, with some secondary stars escaping the gravity
of \sgra. The binary disruption rate is $\sim 10^{-4}$~yr$^{-1}$,
comparable to tidal disruption models.  Stellar ejection rates depend
on the degree of binary hardening, which may occur during migration in
the gaseous disc from which the binary stars formed.  Three cases - no
hardening, moderate hardening (binary separation reduced by a factor
2) and strong hardening (reduction by 10) give ejection rates of $4$,
$7$ and $22 \times 10^{-6}$~yr, respectively. Of these stars, a
fraction would have terminal velocities large enough to be classed as
HVSs observationally. For the three models, we estimate HVS
populations of $37$, $75$ and $512$ stars respectively, although these
numbers are certainly optimistic.

Our models also preserve the spatial anisotropy of the original
stellar population. However, stronger binary hardening increases the
HVS production rate at the expense of dampening the spatial
anisotropy.  Taking these factors together requires massive binaries
formed in the few central parsecs of the Galactic centre to have had
typical separations a factor 2 less than those in the field at the
point just prior to the supernova explosion. This constraint would be
significantly weakened if the hardening process also caused the binary
orbital planes to become somewhat correlated with the plane of the
stellar disc, thus helping to preserve the spatial anisotropy of the
ejected HVSs. We note that the similar HVS production rates predicted
by the SN-induced and tidal disruption models mean that both
mechanisms likely contribute to the observed HVS population.

\vspace{1.0cm}
\noindent
We thank Sergei Nayakshin and Richard Alexander for helpful discussions. 
Astrophysics research at the University of Leicester is supported by the STFC.

\end{document}